\documentclass[twocolumn, twocolappendix]{aastex631}

\usepackage{amsmath}
\usepackage{amssymb}
\usepackage{bm}
\usepackage{xcolor}

\received{August 6, 2024}
\revised{September 23, 2024}
\accepted{October 1, 2024}

\newcommand{\parens}[1]{\left(#1\right)}
\newcommand{\brackets}[1]{\left[#1\right]}
\newcommand{\cxo}{\textit{CXO}}

\submitjournal{ApJ}

\graphicspath{{./}{figs/}}
\begin{document}

\title{A Catalog of Pulsar X-ray Filaments}

\author[0000-0002-6401-778X]{Jack T. Dinsmore}
\affiliation{Department of Physics, Stanford University, Stanford CA 94305}
\affiliation{Kavli Institute for Particle Astrophysics and Cosmology, Stanford University, Stanford CA 94305}

\author[0000-0002-6401-778X]{Roger W. Romani}
\affiliation{Department of Physics, Stanford University, Stanford CA 94305}
\affiliation{Kavli Institute for Particle Astrophysics and Cosmology, Stanford University, Stanford CA 94305}

\begin{abstract}
We present the first \textit{Chandra} X-ray Observatory (\cxo) catalog of ``pulsar X-ray filaments,'' or ``misaligned outflows.'' These are linear, synchrotron radiating features powered by ultra-relativistic electrons and positrons that escape from bow shock pulsars. The filaments are misaligned with the (large) pulsar velocity, distinguishing them from the pulsar wind nebula (PWN) trail which is also often visible in \cxo\ ACIS images. Spectral fits and morphological properties are extracted for five secure filaments and three candidates using a uniform method.  We present a search of archival \cxo\ data for linear diffuse features; the known examples are recovered and a few additional weak candidates are identified. We also report on a snapshot \cxo\ ACIS survey of pulsars with properties similar to the filament producers, finding no new filaments, but some diffuse emission including one PWN trail. Finally, we provide an updated model for the pulsar properties required to create filaments in light of these new observations.
\end{abstract}

\section{Introduction} \label{sec:intro}
Five pulsars have been observed to sport extended X-ray ``filaments'' with several unusual features. They extend for $\sim$ pc into the interstellar medium, retaining a narrow and linear profile. The direction is oblique to the often-large pulsar proper motion, meaning they cannot represent a pulsar wind nebula (PWN) supersonic ``trail.'' Instead, the filaments are thought to consist of synchrotron-radiating, high-energy positrons and electrons escaping to the ambient interstellar medium (ISM) near the pulsar bow shock. Their streaming along ISM magnetic field lines gives them their narrow, linear shape. In some cases a faint antipodal ``anti-filament'' is seen, though in general the structures occur on one side of the pulsar only. The pulsar properties required for particle escape and the details of the particles' transport through the ISM are not yet fully understood. 

The first filament was observed issuing from PSR B2224+65 \citep{cordes1993guitar} in the Guitar nebula. Filaments were also revealed in PSR J2030+4415 \citep{de2020psr, de2022long} and PSR J1101$-$6101 \citep{pavan2014long,klingler2023nustar}; in both cases the filament is at large angle to a clear PWN trail. Two more pulsars, PSR J2055+2539 \citep{marelli2016tale, marelli2019two} and PSR J1509$-$5850 \citep{hui2007radio, klingler2016chandra} also show two outflows, though it is not immediately clear which is the filament and which the trail.

Spectral studies of all these filaments reveal hard power-law spectra---synchrotron radiation from ultra-relativistic electrons and positrons escaping through the pulsar bow shock. They imply equipartition magnetic field values that are generally amplified above the ISM value of $3-5$ $\mu$G. However, the lifetime of X-ray emitting particles in these fields is $\sim$centuries---too long to explain the short filament length if the particles are propagating ballistically. \cite{olmi2024nature} propose to solve this puzzle by amplifying the filament magnetic field far above the equipartition value. Many amplifying processes would cause particle scattering and broaden the filament. This model seeks to avoid scattering by generating turbulence at scales smaller than the particle Larmor radius, although that may not adequately accelerate cooling.

A separate puzzle is how particles are ejected from the bow shock in the first place. \cite{bandiera2008on} highlighted the key role of the stand-off distance
\begin{equation}
  r_0 = \sqrt{\frac{\dot E}{4\pi c \mu m_p n_\mathrm{ISM} v_\mathrm{psr}^2}},
  \label{eqn:r0}
\end{equation}
which is the distance at which the pulsar wind pressure matches the ram pressure of the pulsar through the ISM. Here, $\dot E$ is the pulsar spin-down power, $\mu \approx 1.38$ is the ISM mean molecular mass, $m_p$ is the proton mass, $n_\mathrm{ISM}$ is the ISM number density, and $v_\mathrm{psr}$ is the pulsar velocity. A particle with Larmor radius $r_L \gtrsim r_0$ may escape the bow shock to create a filament, while lower energy particles are trapped inside. This should only occur for the highest velocity pulsars in the densest parts of the ISM. \citet{2019MNRAS.490.3608O,2019MNRAS.484.5755O} have generated MHD simulations which explore these ideas (see \cite{2023Univ....9..402O} for a review).

This paper provides the first catalog of known pulsar X-ray filaments, suitable for validating these and other theories regarding particle escape from the bow shock and propagation within filaments. In section \ref{sec:population}, we tabulate the properties of the five known pulsar filaments and suggest three additional filament candidates. We further analyze the observed spectrum (section \ref{sec:spectrum}) and morphology (section \ref{sec:morphology}) and discuss commonalities between the observed filaments. Section \ref{sec:search} searches for new filaments in both archival data (section \ref{sec:archival}) and recent observations by the \textit{Chandra} X-ray Observatory (\cxo) (section \ref{sec:new}). We provide a heuristic condition for particle escape, calibrated on these results, in section \ref{sec:escape}. We summarize these results in Section \ref{sec:conclusion}.
\section{The catalog} \label{sec:population}
\begin{figure}
  \centering
  \includegraphics[width=\linewidth]{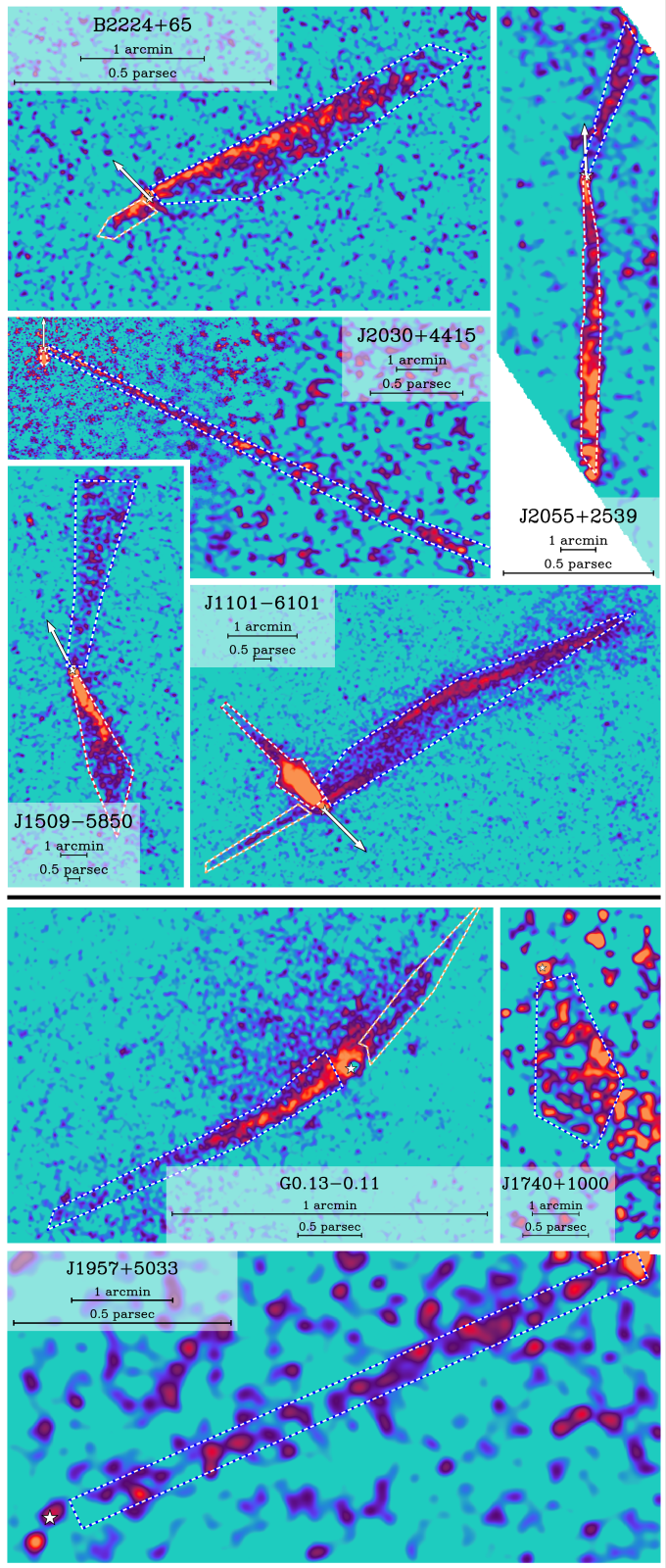}
  \caption{Stacked, exposure-corrected \cxo\ observations of known (top) and candidate (bottom) filaments. Regions used for spectral fitting for filaments, anti-filaments and trails are marked with dotted lines. Bright point sources have been removed. The pulsar position is marked with a star and the proper motion indicated if known or assumed. Celestial north points up.}
  \label{fig:filaments}
\end{figure}

The catalog of pulsar X-ray filaments currently contains five objects (Fig.~\ref{fig:filaments}), which satisfy the defining properties of being extended, hard, linear X-ray sources associated with pulsars and misaligned with the pulsar proper motion. We also discuss three candidates, which fail to match at least one of these criteria.

PSR B2224$+$65 (Guitar), the prototypical and first discovered, is a bright and short X-ray filament. The pulsar exhibits an unusually large VLBI-measured proper motion and an H$\alpha$ bow shock \citep{cordes1993guitar}. It has a moderate anti-filament, and a very dim PWN trail \citep{2021RNAAS...5....5W}.

PSR J2030$+$4415 has a very long filament of low surface brightness, so much of its X-ray flux is visible as a PWN trail \citep{de2022long}. It has an X-ray measured proper motion and an H$\alpha$ bow shock. Like most filament pulsars discovered so far it is radio-quiet and so no dispersion measure (DM)-based distance estimate is available. The latest $\gamma$-ray flux \citep{2023ApJ...958..191S} gives a heuristic estimated $d=0.89$\,kpc, while estimates from the H$\alpha$ velocity spread give $\sim 0.5-0.8$\,kpc \citep{de2020psr}; here we adopt $d= 0.75\pm0.15$\,kpc and hence $v_\perp \approx 300$\,km/s.

PSR J1101$-$6101 (Lighthouse) is the brightest and most distant X-ray filament \citep{pavan2014long}, boasting an accompanying PWN trail and the brightest anti-filament in the catalog. An analysis from NuSTAR data observed a cooling break along its length \citep{klingler2023nustar} in photons up to 25\,keV. No other filament shows a similarly clear break. Though no proper motion has been measured, Lighthouse's association with a 10--30 kyr old supernova remnant (SNR) at a projected distance of $\sim$18\,pc \citep{garcia2012origin} implies a very large space velocity. We adopt $v_\perp\sim900$\,km s$^{-1}$ for $t=20$\,kyr. The pulsar itself exhibits a larger characteristic age of 116\,kyr, implying that it was born close to its present 63\,ms spin period.

\begin{table*}
  \movetableright=-1cm
  \begin{tabular}{l|ccccc|ccc}
    \hline\hline
    \textbf{Pulsar} & $\dot E$ [erg s$^{-1}$]  & $\tau_c$ [kyr] & $d$ [kpc] & $v_\perp$ [km s$^{-1}$]  & $|z|$ [pc] & $r_0$ [AU] & $\gamma_\mathrm{MPD}$ & $g$ \\\hline
    J1101--6101 & $1.4\times 10^{36}$ & 116 &  $7^\mathrm{c}$& $1800 - 600^\mathrm{g}$  & 114 & 1291 & $4.0\times 10^{9}$ & 94\\ 
    B2224+65 & $1.2\times 10^{33}$ & 1120 &  $0.83^\mathrm{a}$ & $760^\mathrm{a}$  & 99 & 46 & $1.2\times 10^{8}$ & 86\\ 
    J2030+4415 & $2.2\times 10^{34}$ & 555 &  $0.75^\mathrm{b}$& $30 0^\mathrm{b}$  & 38 & 494 & $5.0\times 10^{8}$ & 36\\ 
    J1509--5850 & $5.1\times 10^{35}$ & 154 & $3.8^\mathrm{d}$& --  & 41 & -- & $2.4\times 10^{9}$ & --\\ 
    J2055+2539 & $5.0\times 10^{33}$ & 1230 &  $0.4^\mathrm{e}$& --  & 87 & -- & $2.4\times 10^{8}$ & --\\  \hline
    G0.13--0.11 & -- & -- & $8.5^\mathrm{f}$& --  & 16 & -- & -- & --\\ 
    J1740+1000 & $2.3\times 10^{35}$ & 114 &  $1.2^\mathrm{d}$& --  & 425 & -- & $1.6\times 10^{9}$ & --\\ 
    J1957+5033 & $5.3\times 10^{33}$ & 838 &  $0.9^\mathrm{e}$& --  & 172 & -- & $2.5\times 10^{8}$ & --\\ 
    
    \hline\hline
    \end{tabular}
    
  \caption{Properties of the pulsars that host filaments. Directly observed properties are shown on the left while inferred properties are at right. Columns give spin-down luminosity, characteristic age $\tau_c=\dot P / (2P)$, distance, velocity transverse to the line of sight, height above the Galactic disk, stand-off distance, Lorentz factor corresponding to the pulsar maximum potential drop, and the ``gate fraction'' $g$ which may control filament particle escape (Eq.~\ref{eqn:gate-fraction}). $\dot E$ and coordinates are retrieved from the ATNF catalog \citep{manchester2005australia}. The table is ranked by $g$. Other references are $^\mathrm{a}$\citep[parallax]{deller2019microarcsecond}, $^\mathrm{b}$see text, $^\mathrm{c}$\citep[associated SNR velocity]{reynoso2006high}, $^\mathrm{d}$\citep[DM]{yao2017new}, $^\mathrm{e}$\citep[$\gamma$-ray heuristic]{parkinson2010eight}, $^\mathrm{f}$(Distance to Galactic center), $^\mathrm{g}$\citep{garcia2012origin}.}
  \label{tab:pulsar}
\end{table*}

PSR J1509$-$5850 shows two linear structures, with the southern system exhibiting a swept-back compact nebula that identifies it as the likely trail \citep{klingler2016chandra}. That trail, like other PWN sources, also exhibits extended radio emission \citep{hui2007radio} though the pulsar itself and the northern filament are radio quiet. The trail hosts a faint H$\alpha$ shock structure at its apex \citep{brownsberger2014survey}.

PSR J2055$+$2539 is another double-trail pulsar, showing a filament and a PWN trail. We suggest that the northern branch is the filament, due to its harder spectrum (once coincident stars are masked), though the opposite assignment has also been suggested \citep{marelli2016tale}.

The candidate filament G0.13$-$0.11 is a well-observed, two-sided, extended structure near the Galactic center \citep{johnson2009large}. It emits in radio, X-rays, and gamma-rays, as observed by a variety of spacecraft and the ground-based H.E.S.S air Cherenkov telescope. It lies at the cusp of (and may power) the radio arc. The Imaging X-ray Polarimetry Explorer (\textit{IXPE}) has detected highly polarized synchrotron emission indicating a magnetic field aligned with the extended X-ray structure, which we treat as the filament \citep{churazov2024pulsar}. The point source is suspected to be a pulsar \citep{2002ApJ...581.1148W} but no pulsar is as yet identified.

The proper motion of the candidate filament PSR J1740$+$1000 has not been measured, but the pulsar is at high-Galactic latitude ($|z|= 425$\,pc) and its young spin-down age of 114\,kyr \citep{mclaughlin2002psr} implies a high proper motion if the pulsar was born in the disk. Analysis of XMM data showed a hard power-law spectrum \citep{kargaltsev2008x}, though without a detected, misaligned proper motion we cannot exclude the possibility that the object is simply a PWN trail. Morphologically, the candidate is not as narrow as the confirmed filaments.

Finally, the dim candidate filament PSR J1957$+$5033 has been observed with XMM-Newton and \cxo\ and exhibits a long, linear nebula \citep{zyuzin2022likely}, revealed to be slightly curved in XMM data. Like J1740, J1957 is also at high Galactic latitude ($|z|= 153$ pc) and the filament candidate extends away from the disk. If the filament were a PWN trail, the pulsar must have originated far from the disk. While the candidate's morphology is similar to J2030's, the (poorly measured) spectral index is softer than that of most filaments.

We focus on the \cxo\ observations of these filaments in this work, since these structures are narrow and generally are only resolved with \textit{Chandra}. Nevertheless all except J2030 have also been observed with XMM-Newton. In addition, all the pulsars are associated with point sources in the Fermi DR4 catalog \citep{abdollahi2020fermi} except Guitar, though \cite{klingler2023nustar} point out that 4FGL J1102.0-6054 may instead originate from a nearby blazar.

\subsection{Pulsar Properties}
We present relevant observed properties of these host pulsars in table \ref{tab:pulsar}. Stand-off distance $r_0$ was computed assuming $\mu = 1.38$ and $n_\mathrm{ISM} = 0.5$ cm$^{-3}$ and does not correct the observed $v_\perp$ to $v_{\rm psr}$, since the 3-D space velocity is not generally known. Dashes represent unmeasured quantities. While we do not pursue a full population analysis of these properties, we provide some context in the remainder of this section.

The \cite{bandiera2008on} picture of small $r_0$ giving rise to filaments requires that transverse velocity $v_\perp \propto v_{\rm PSR}$ be large and spin-down luminosity $\dot E$ small. Velocity measurements, when available, support this; the lowest transverse velocity (J2030's 300\,km s$^{-1}$) is in the 84th percentile of the known velocities in the ATNF\footnote{\url{http://www.atnf.csiro.au/research/pulsar/psrcat/}} catalog \citep{manchester2005australia}. The next smallest (Guitar's 760 km s$^{-1}$) is above the 95th percentile. However, the large variance in spin-down luminosities disagrees with the naive expectation from simple $r_0$ selection. All filaments except Guitar are more luminous than the median ATNF $\dot E$ of $3.6 \times 10^{33}$\,erg s$^{-1}$. We discuss a potential explanation in section \ref{sec:search}. 

The remaining physical parameter entering Eq.~\ref{eqn:r0} is the poorly constrained ISM number density $n_\mathrm{ISM}$. This quantity varies strongly across the Galaxy through the ISM phases, but is largest when the height above the Galactic disk $|z|$ is low. Our pulsars show a slight preference for low $|z|$; they are closer to the disk than the median ATNF pulsar of $|z| = 390$ pc and more typical of the 25th percentile $|z| = 150$ pc. J2030 is the closest filament pulsar to the disk---closer than 90\% of ATNF pulsars. This may counterbalance its relatively low velocity $v_\mathrm{psr}$. Another indication of large $n_\mathrm{ISM}$ is filaments' unusually common association with H$\alpha$ bow shocks, which require dense, largely neutral ambient ISM to form. Three out of five filaments host these shocks---a sizeable fraction of all known H$\alpha$ bow shock pulsars.

\subsection{Spectral Properties}
\label{sec:spectrum}

\begin{figure*}
  \centering
  \includegraphics[width=0.9\linewidth]{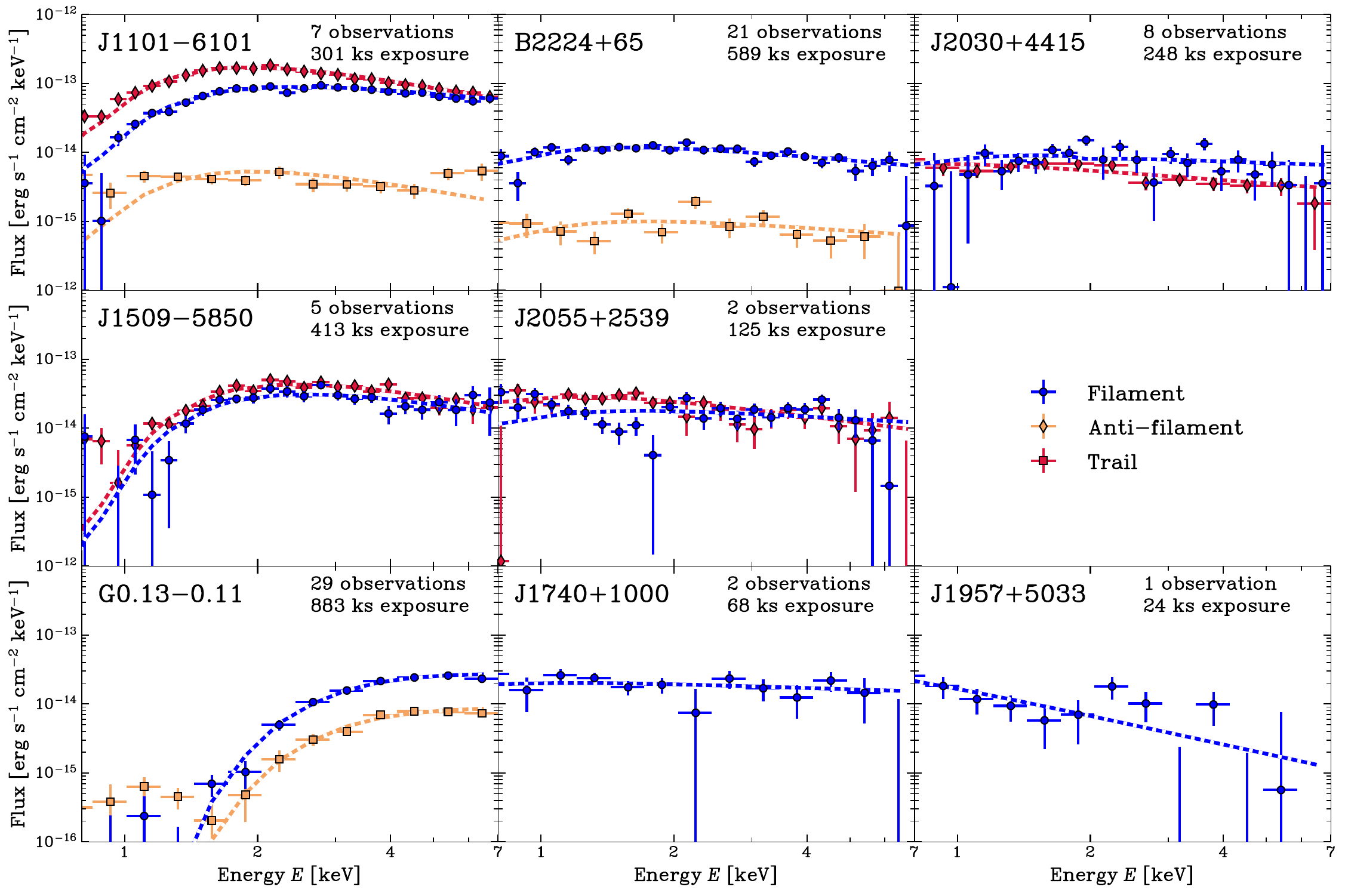}
  \caption{Filament spectra as observed by \cxo\ with best fit absorbed power laws. Though the data shown are
  background-subtracted, we perform the fit on un-subtracted data to preserve signal to noise.}
  \label{fig:spectra}
\end{figure*}

\begin{table*}
  \centering
  \begin{tabular}{l|cccccccc}
    \hline\hline
    \textbf{Pulsar} & Exposure [ks] & $n_H$ [$10^{22}$ cm$^{-2}$] & $\Gamma_\mathrm{fil}$ & $\Gamma_\mathrm{trail}$ & $\Gamma_\mathrm{anti}$ & $L_\mathrm{fil} / 10^{-4}\dot E$ & $L_\mathrm{trail} / L_\mathrm{fil}$ & $L_\mathrm{anti} / L_\mathrm{fil}$ \\\hline
    J1101--6101 & 301 [60] & $0.97$(4) & $1.63$(3) & $2.17$(4) & $2.1$(2) & $30.7(8)$ & $2.1$(1) & $0.07(1)$  \\ 
    B2224+65 & 589 [45] & $0.30$(6) & $1.6$(1) & -- & $1.5$(2) & $49(5)$ & $0.026(6)$ & $0.09(1)$  \\ 
    J2030+4415 & 248 [--] & 0.06$^\mathrm{a}$ & $1.3$(2) & $1.5$(1) & -- & $4.1(5)$ & $0.23$(2) & --  \\ 
    J1509--5850 & 413 [62] & $1.7$(2) & $2.0$(2) & $2.1$(2) & -- & $11(2)$ & $1.5$(3) & --  \\
    J2055+2539 & 125 [123] & 0.22$^\mathrm{b}$ & $1.4$(1) & $1.8$(1) & -- & $5.1(4)$ & $1.21$(9) & --  \\  \hline
    G0.13--0.11 & 883 [15] & $6.1$(2) & $1.2$(1) & -- & $1.2$(2) & -- & $1.1$(1) & $0.31$(5)  \\ 
    J1740+1000 & 68 [449] & 0.05$^\mathrm{c}$ & $1.2$(2) & -- & -- & $1.0(1)$ & -- & --  \\ 
    J1957+5033 & 24 [34] & 0.03$^\mathrm{d}$ & $2.4$(5) & -- & -- & $7(2)$ & -- & --  \\ 
    
    \hline\hline
  \end{tabular}
  
  \caption{Spectral properties of the known (top) and candidate (bottom) filaments. Unabsorbed luminosities are given in the $0.5-7$ keV range. Dashes indicate that the feature is not bright enough for a spectral fit. The exposures in brackets represent on-axis XMM PN exposures which we do not use in our fits. In some cases, hydrogen column density was obtained externally, from $^\mathrm{a}$\citep{de2022long},  $^\mathrm{b}$\citep{marelli2016tale},  $^\mathrm{c}$\citep{kargaltsev2008x},  $^\mathrm{d}$\citep{zyuzin2022likely}. }
  \label{tab:spectral}
\end{table*}

Previous works have shown that filament spectra are well modeled by an absorbed power law $dN/dE \propto E^{-\Gamma}$. We perform this fit to all observed filaments with a consistent method. When present, we also fit a power law to the PWN trail and the anti-filament. Guitar contains a PWN trail but not enough flux is collected to support a spectral index measurement. Regions are shown in Fig.~\ref{fig:filaments}.

Only \cxo\ data are used because the observatory's superior point spread function (PSF) renders it the most reliable for faint source extraction, though other data are available. Observations with aim-point $>3'$ from the filament are excluded due to \cxo's increased off-axis PSF. The data are first reprocessed using the standard \texttt{ciao} tools and point sources are removed with a custom method (appendix \ref{app:star}). We correct for proper motion by aligning the pulsar position in each observation. We simultaneously fit background and source models to all observations, where the background consists of an astrophysical absorbed power law plus a particle distribution obtained from the \cxo\ stowed observations. The normalization
of the particle background is allowed to vary independently for each observation. After binning the $0.5-7$ keV data, we model the uncertainty of each bin as Poisson-distributed and fit by minimizing the Cash statistic. For fields with multiple structures, the atomic hydrogen column density $n_H$ of the dimmer structures is fixed to the best-fit value of the brightest.

Spectra are shown in Fig.~\ref{fig:spectra} and best fit results are reported in table \ref{tab:spectral}. Unabsorbed fluxes in the $0.5-7$ keV band are converted to luminosities using the pulsar distances in table \ref{tab:pulsar} (distance uncertainties not propagated) and are compared to other relevant luminosities (uncertainty of the denominator not propagated). A dash represents that the given feature was not observed.

Excluding the conical J1509, the average filament spectral index is $\Gamma \approx 1.5$, moderately harder than typical PWN indices \cite[see][]{2008AIPC..983..171K}, which exhibit a median PWN index of $\Gamma=1.7$. Indeed, every filament exhibits a harder filament index than its PWN trail, if one exists. The candidate G0.13 shows an excess of X-ray emission behind the filament and anti-filament, likely representing particles escaped from the filament rather than a typical PWN trail. Its spectral index is comparable to that of the filament and anti-filament.

The $L_\mathrm{fil}/\dot E$ ratio---i.e. the efficiency of spin-down power conversion to filament luminosity---varies dramatically among the filaments. Guitar boasts the largest ratio, and interestingly the smallest stand-off distance. However, the second largest ratio is emitted by Lighthouse, which possesses a large $r_0$ due to its much larger $\dot E$. The cause of $L_\mathrm{fil}/\dot E$ variation is unclear and may stem from multiple physical parameters.

Anti-filaments are observed in three of the four deepest observations available, with J1509 being the one deep observation without an anti-filament. However, J1509's bright trail may obscure a dim anti-filament if present. J2030 has the next-deepest observation but shows no anti-filament. There may be some correlation between filament and anti-filament spectral properties; in Guitar and G0.13, the filament and anti-filament spectral indices are similar, while in Guitar and Lighthouse the filament and anti-filament luminosity ratios agree. These trends are not universal; Lighthouse's anti-filament is softer than its filament and G0.13's anti-filament is closer in flux to its filament than for Lighthouse or Guitar. More anti-filament examples are necessary to delineate their properties. 

The $L_\mathrm{trail}/L_\mathrm{fil}$ ratio varies greatly with no clear trend, in contrast to the strong $\Gamma_\mathrm{trail} > \Gamma_\mathrm{fil}$ trend. Certainly trails and filaments have different particle populations and cooling behavior. Pulsar spin and magnetic geometry might be important for the trail/filament partition.

The limited \cxo\ band does not allow us to probe whether the filament spectrum continues to lower energies. However, the lack of radio detections of any pulsar X-ray filament implies a strong low energy cutoff in the filament particle distribution. The fact that many PWN trails are detected in the radio, even for filament pulsars, suggests that the cut-off is imposed when particles are injected into the filament, with lower energy particles swept back in the trail.

\subsection{Morphological Properties}
\label{sec:morphology}

\begin{table*}
  \centering
  \begin{tabular}{l|ccccc|ccc|c}
    \hline\hline
    \textbf{Pulsar} & $\ell$ [pc] & $w$ [pc] & $R$ [pc]  & $\theta_\mathrm{trail}$ [$^\circ$]  & $\theta_\mathrm{anti}$ [$^\circ$] & $c \tau_\mathrm{cool}$ [pc] & $v_\perp \tau_\mathrm{cool}$ [pc] & $r_{L,\mathrm{MPD}}$ [pc] & $B$ [$\mu$G]\\\hline
    J1101--6101 & 9.4 & 1.13 & $22.$ & 100 & 178 & $430$  & $1.3$ & $0.26$ & $8.6$\\ 
    B2224+65 & 0.6 & 0.06 & -- & 64 & 172 & $200$  & $0.51$ & $0.0045$ & $14$ \\ 
    J2030+4415 & 3.5 & 0.02 & $25.$ & 64 & -- & $130$  & $0.13$ & $0.015$ & $19$ \\ 
    J1509--5850 & 8.5 & 0.88 & -- & 139 & -- & $540$  & -- & $0.18$ & $7.3$\\
    J2055+2539 & 0.5 & 0.03 & -- & 159 & -- & $180$  & -- & $0.0086$ & $15$ \\  \hline
    G0.13--0.11 & 2.4 & 0.19 & $32.$ & -- & 155 & $90$  & -- & -- & $24$ \\ 
    J1740+1000 & 0.9 & 0.28 & -- & -- & -- & $640$  & -- & $0.14$ & $6.6$\\ 
    J1957+5033 & 3.6 & 0.03 & $7.1$ & -- & -- & $130$  & -- & $0.0072$ & $19$ \\ 
    
    \hline\hline
  \end{tabular}

  \caption{Morphological properties of the known filaments. The length $\ell$, width $w$, radius of curvature $R$, filament-trail angle $\theta_\mathrm{trail}$, and filament-anti-filament angle $\theta_\mathrm{anti}$ (\textit{left}) are measured. The potentially relevant physical scales (\textit{middle}) are computed using equipartition magnetic fields (\textit{right}).}
  \label{tab:morphology}
\end{table*}

A morphological model for filaments would shed light on the manner of particle transport and confinement that the spectrum cannot address. Lacking a full model for filament morphology, we report measurements of the filament characteristic scales, such as their length, width and radius of curvature. We compare these measurements to three important physical scales
\begin{enumerate}
  \item The cooling length $c\tau_\mathrm{cool}$, which is the distance a relativistic particle travels before cooling from synchrotron radiation. This is a natural scale for the filament length.
  \item The distance $v_\perp\tau_\mathrm{cool} \propto v_\mathrm{psr}\tau_\mathrm{cool}$ that the pulsar travels during the lifetime of a single particle. This is a natural scale for the filament width.
  \item The maximum Larmor radius $r_\mathrm{L,MPD} = \gamma_\mathrm{MPD} mc^2 / (eB_\mathrm{eq})$ of a particle produced by the pulsar ($\gamma_\mathrm{MPD}$ defined in Eq.~\ref{eqn:gamma-mpd}). This is a natural maximum for the sharpness of the filament leading edge.
\end{enumerate}
These scales rely on the timescale for an electron to shed its energy by synchrotron cooling
\begin{equation}
  \tau_\mathrm{cool} = 49.7\  B_\mathrm{\mu G}^{-3/2} E_\mathrm{peak}^{-1/2}\ \mathrm{kyr}.
\end{equation}
where $E_\mathrm{peak}$ is the peak photon energy emitted by the particle in keV. We take $E_\mathrm{peak}=2$, typical for the \cxo\ effective area. $\tau_\mathrm{cool}$ is calculable under the assumption of equipartition magnetic fields
\begin{equation}
  B_\mathrm{eq} = 3.7\brackets{\frac{L_\mathrm{32}}{\phi V_\mathrm{pc}} \frac{C_{1.5 - \Gamma}(E_\mathrm{min}, E_\mathrm{max})}{C_{2 - \Gamma}(\epsilon_\mathrm{min}, \epsilon_\mathrm{max})} X(\Gamma)}^{2/7}\mu \mathrm{G},
  \label{eqn:equipartition}
\end{equation}
where $C_q(a, b) = (b^q - a^q) / q$, $L_{32}$ is the filament luminosity in units of $10^{32}$\,erg\,s$^{-1}$, $V_\mathrm{pc}$ is the emitting volume in pc$^{3}$, $\phi$ is the volume filling factor of the emitting particles, $E_\mathrm{max/min}$ represent the peak radiation emitted by the most- and least-energetic particle in units of keV, and $\epsilon_{\mathrm{max/min}}$ represent the \cxo\ band ($0.5-7$) in keV. $X(\Gamma)$ corresponds to the numerical factor
\begin{equation}
  \begin{aligned}
    X(\Gamma) = \brackets{6.9^{\Gamma - 2}\Gamma_f\parens{\frac{\Gamma}{2}-\frac{1}{3}}\Gamma_f\parens{\frac{\Gamma}{2}+\frac{4}{3}}}^{-1}\Gamma
  \end{aligned}
\end{equation}
where $\Gamma_f$ is the gamma function (Appendix \ref{app:equipartition}). We take $\phi=1$ and $V_\mathrm{pc} = (\pi/4) \ell w^2$ (a cylinder), where $\ell$ is the observed length of the filament and $w$ is the observed width.

The projected filament width $w$ and radius of curvature $R$ are obtained from a fit to \cxo\ data by the filament search method described in the next section, which corrects for systematics such as exposure and energy sensitivity. $R$ could not be measured for the shortest and widest filaments because the model over-fits to internal structure. Angular distances are measured and converted to physical distances using the assumed pulsar distance $d$ in table \ref{tab:pulsar}. All filaments are wider than the \cxo\ PSF, though J2030 is only barely resolved.

Projected length is measured by hand from the pulsar to the farthest point where the filament is still clearly discernable above the background. Two filaments, J2055 and J1957, extend outside the \cxo\ field of view. XMM observations do not show any further extension to J2055, though J1957 may extend two to three times farther than the \cxo\ image suggests. Lengths measured from XMM data may underestimate the true length, because filament widths are unresolved by XMM and may be dominated by background near the faint end.

These lengths and widths are used to estimate the equipartition magnetic field, with the exception of J2030 for which we report equipartition magnetic fields from the first $\sim 280''$ where exposure is deep. Using the observed $\ell$ underestimates the true length (and therefore overestimates $B$) due to projection effects, while assuming $\phi=1$ underestimates $B$. Further discussion of Eq.~\ref{eqn:equipartition} is given in appendix \ref{app:equipartition}.

Morphological results are given in table \ref{tab:morphology}. The equipartition magnetic fields strengths are elevated slightly above typical the ISM 3--5 $\mu$G field for all filaments. Nevertheless, the magnetic fields are still too weak to cool particles by the time they reach the end of the filament ($\ell \ll c\tau_\mathrm{cool}$). The filament length can still be explained if particles are trapped and do not move ballistically. Alternatively, the magnetic field could be amplified far above equipartition, but this is difficult to accomplish. To affect synchrotron power at X-ray wavelengths, the magnetic field must be amplified at scales comparable to the Larmor radius, in contrast to the assumptions of \citet{olmi2024nature}. But strong amplification at these scales would also scatter particles, broadening the filament.

The cooling scale associated with pulsar motion is much closer to the observed filament width, though some magnetic field amplification is required for an exact match. This suggests a model in which a one-dimensional filament is created by the pulsar and remains stationary with respect to the ISM, cooling in the wake of the pulsar in an amplified magnetic field until the highest energy particles fall out of the \cxo\ band and the filament fades.

Finally, the Larmor radius of the most energetic particles is generally smaller than the width by a factor of $\sim 10$. This sets a natural length scale for intrinsic ``blurring'' effects due to motion of filament particles, such as a potential dulling of the filament's sharp leading edge.

The radius of curvature $R$ is likely unconnected to these length scales, instead set by the scale of variations in the ISM magnetic field. For J1957, the fit is performed on the \cxo-observed first 1.5 pc of the filament. Though the more distant XMM data is visibly curved, the curvature occurs in the opposite direction, making an $S$ shape. These lengths are indeed comparable to the 10$-$100 pc ISM correlation length, e.g. as determined from Faraday rotation differences between pulsars \citep{ohno1993random}.

The trail-filament angles appear slightly biased towards antiparallel, but this is not yet a significant trend. The filament-anti-filament angle is by definition $<$180$^\circ$; the substantial decrease in two cases could be the effect of the magnetic field draping around the pulsar.

\section{Searching for Additional Filaments} \label{sec:search}

The previous sections show an evident trend that filament pulsars have large velocity and small $r_0$. However, additional factors clearly control filament presence and brightness, including the local ISM density and magnetic field and plausibly the pulsar spin and magnetic field geometry \citep[cf.][]{de2022long}. We would clearly like to discover and measure more examples to probe such effects.

This proves to be challenging. Due to their low luminosity, some known X-ray filaments have historically been overlooked until follow-up observations of their host pulsars were made (e.g.~J2030). Thus we first analyze the archive of \cxo-observed pulsars searching for unremarked linear structures. Results show that the \cxo\ archive contains no decisive filaments beyond those contained in this work. Evidence for non-background nebulosity was uncovered for three pulsars. In addition we report on a snapshot survey of unobserved pulsars whose properties suggest that they might host filaments. 

\subsection{Archival Search Methods}
\label{sec:search-methods}

\begin{figure*}
  \centering
  \includegraphics[width=\linewidth]{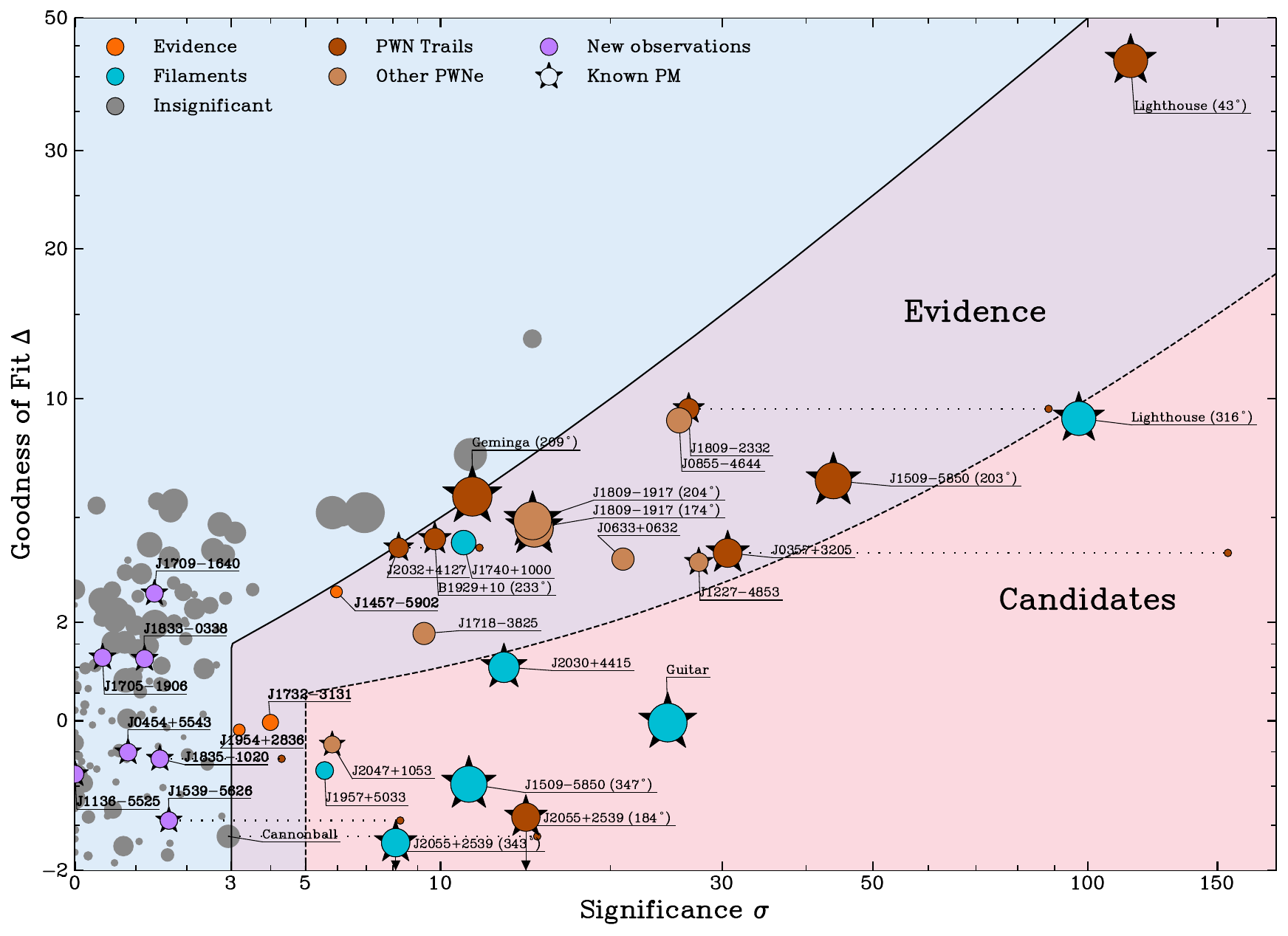}
  \caption{Results of a search for filament candidates among \cxo\ observations. The pink and purple background represent strong / weak acceptance regions for filament candidates; low $\Delta$ and high $\sigma$ are preferred. Blue points represent known filaments or candidates listed in this work, brown represent known non-filament PWNe, and the three orange points are filament candidates caught by the search. Radii scale logarithmically with total exposure time. For trails, a small circle shows the elevated significance arising from anti-alignment with the proper motion. Axes are shifted-logarithmic to better display data with $\sigma$ or $\Delta$ near zero. An online version together with images of each pulsar is available at \url{https://jack-dinsmore.github.io/filament-website/filament.html} (Permanently hosted at \cite{jack_dinsmore_2024_13830862}).
  }
  \label{fig:search-results}
\end{figure*}

Our pulsar sample consists of all \cxo\ ACIS observations with aim-point within 2$'$ of a pulsar in the ATNF catalog. To avoid bright extended structures such as SNRs, we further require spin-down age $\tau > 50$ kyr if known, and no coincident sources in the \cite{green2019revised} catalog of SNRs. Continuous clocking observations are excluded. A small fraction of observations are polluted by nearby source complexes, such as globular clusters, or by image systematics, such as bright readout streaks. We manually remove these. The remaining 158 pulsars (291 observations; 36 pulsars have multiple observations) are processed with the following algorithm. 

The data are reprocessed using the \texttt{ciao} tools and point sources are removed using a custom method (appendix \ref{app:star}). The remaining $0.7-6$ keV data within 20$''$ to 4$'$ of the source are binned into $B$ bins of a few square arcseconds each, and a simple filament model is fitted using the likelihood 
\begin{equation}
  \mathcal{L} = \prod_{b=1}^B P(n_b|\lambda_+ + \lambda_-) \prod_{i=1}^{n_b} \frac{\lambda_+ P_+(E_i) + \lambda_- P_-(E_i)}{\lambda_+ + \lambda_-}.
  \label{eqn:search-likelihood}
\end{equation}
Here, $n_b$ are the number of counts in bin $b$, $\lambda_+$ and $\lambda_-$ are the expected number of counts from the filament and the constant-flux background respectively in bin $b$, and $E_i$ are the energies of the events detected within the bin. $P(n_b | \lambda)$ is the Poisson probability of detecting the observed number of counts, while $P_+$ and $P_-$ are the probabilities of detecting the observed energies from the filament or background (proportional to the spectra).

For this search, we set the filament profile to be a uniform-flux box with variable flux and width. Both the background and filament expected counts are weighted by the \texttt{ciao}-generated exposure map.

The filament spectrum is taken to be an absorbed power law with $\Gamma=1.6$ and $n_H$ given by the $n_H$-DM correlation identified in \cite{he2013correlation}, when a DM is available. Otherwise, the total-column Galactic $n_H$ is obtained from the \texttt{HEAsoft} $n_H$ calculator. The photoelectric absorption cross sections and chemical abundances are obtained from \cite{verner1996atomic} and \cite{anders1989abundances} (the \texttt{phabs} \texttt{XSPEC} model). The instrument energy response function and quantum efficiency is approximated by a weighted ARF computed by the \texttt{ciao} tool \texttt{mkwarf} in a small region at the center of each ACIS chip. This method accounts for the large variations in effective energy between chips and over \cxo's lifetime, but not the small variations across chips.

The fit results are summarized in two statistics: the significance of detection $\sigma$ and goodness of fit (GoF) $\Delta$. Significance is defined in the standard way: it is a function of the log-likelihood ratio $\sigma = f_1(\ln \mathcal {L}_M / \mathcal {L}_{NH})$ where $\mathcal{L}_M$ is the best fit likelihood and $\mathcal {L}_{NH}$ is the likelihood of the no-filament null hypothesis. The function $f_1$ is chosen such that $\sigma$ follows the standard Gaussian distribution under the null hypothesis. The GoF is a function of the best fit log-likelihood $\Delta = f_2(\ln \mathcal{L}_M)$ such that $\Delta$ is distributed according to the standard Gaussian under the model. It is analogous to a $\chi^2$ value that has been mapped to a Gaussian distribution. The functions for $\Delta$ and $\sigma$ are derived in appendix \ref{app:search}.

A true filament exhibits high $\sigma$ and low $\Delta$ (with negative $\Delta$ arising from favorable statistical fluctuations). Low $\sigma$ observations are consistent with noise, and objects with high $\sigma$ and high $\Delta$ are non-filament extended structures.

\subsection{Archival Search Results}
\label{sec:archival}
The results of our archival search are shown in Fig.~\ref{fig:search-results}. Each point represents a pulsar. If a pulsar shows two linear structures (one likely a PWN trail), we label these by the position angle measured from North to East. 
We only consider pulsars lying within the ``evidence region,'' defined by $\sigma > 3$ and $\sigma/\Delta > 2$. The second cut removes extended sources that differ strongly from our filament model relative to their signal to noise. These generous cuts were chosen to accept all known filaments, candidates, and filament-like PWN trails. A stricter ``candidate region'' defined by $\sigma > 5$ and $\sigma/\Delta > 10$, which accepts all known filaments, is also drawn. J1740 does not fall in our candidate region because of its wide, conical morphology. We nevertheless promote it due to its significance and hard spectral index.

Many of the low-$\Delta$ structures lie $\sim 180^\circ$ from the observed proper motion, and we label these as PWN trails (dark brown). Trails are often morphologically indistinguishable from filaments. Together, these trails and the known filaments (blue) constitute the \cxo-observed objects that best fit our filament model. Some true trails and filaments yield large $\Delta$, though these are generally wedge-shaped (J1509), curved (Lighthouse), short (Guitar), or otherwise disagree with our simple filament model. $\Delta$ can be improved in these cases by adding parameters to model more complex morphology, but for the sake of a uniform search we do not show these results in Fig.~\ref{fig:search-results}. The light brown points, largely with high $\Delta$, are visually identified to be extended structures such as PWNe. 

Three remaining points (orange) display evidence of non-background nebulosity. The J1457$-$5902 extended emission is relatively bright but wide, morphologically akin to J1740. J1732$-$3131 is narrower (though very dim) and extends for $\sim 3'$. J1954+2836 presents similarly thin but even dimmer morphology. J1954 and J1457 each were observed with a 10 ks exposure and J1732 with 20 ks. With low significance and goodness of fit, these are unlikely to be true filaments. Nevertheless, deeper \cxo\ observations to probe these PWN structures would be valuable.

Unsurprisingly, visual inspections by the original observers have identified all of the ``obvious'' filament candidates. However our survey allows us to quantify the significance of these objects, to select marginal cases that are not visually obvious candidates and, most importantly, to limit the presence of such candidates in other observations. In sum, except for the known filaments and these three pulsars, we exclude the presence of significant filaments in previous \cxo\ exposures of $\tau > 50$\,ky pulsars.

\begin{table*}
  \centering
  \begin{tabular}{c|cccccc|ccc}
    \hline\hline
    \textbf{Pulsar} & $\dot E$ & $\tau_c$ & $d$ & $v_\perp$ & $g$ & $r_0$ & Significance ($\sigma$) & $F_\mathrm{PS}$  & $F_\mathrm{ext}$  \\
    & [10$^{33}$ erg s$^{-1}$] & [kyr] & $d$ [kpc] & [km s$^{-1}$] &  &  [AU] &  &  &   \\ \hline
    J1539$-$5626 & 13  & 795 & 3.54$^\mathrm{a}$ & 1019$^\mathrm{d}$& 53 & 251
      & 1.6$^\dagger$ (8.2)
      & $1.2^{+3.6}_{-1.2}$
      & $36^{+28}_{-27}$ T, $27^{+22}_{-21}$ F \\
    J1835$-$1020 & 8.4  & 810 & 2.91$^\mathrm{a}$ & 349$^\mathrm{e}$ & 18 & 590
      & 1.5$^\dagger$ (4.4)
      & $<1.9$
      & $<34$ T \\
    J0454$+$5543 & 2.4  & 2280 & 1.18$^\mathrm{b}$ & 314$^\mathrm{b}$ & 15 & 350
      & 0.9
      & $1.8^{+3.2}_{-1.4}$
      & $<12$ N \\
    J1136$-$5525 & 6.7  & 702 & 1.52$^\mathrm{a}$ & 318$^\mathrm{d}$ & 15 & 579
      & $<0$
      & $3.9^{+4.35}_{-2.5}$
      & $<19$ N \\
    J1705$-$1906 & 6.1  & 1140 & 0.75$^\mathrm{a}$ & 329$^\mathrm{d}$ & 15 & 534
      & 0.4
      & $<1.5$
      & $13^{+7}_{-5}$ N \\
    J1833$-$0338 & 5.1  & 262 & 2.50$^\mathrm{c}$ & 250$^\mathrm{c}$ & 13 & 642
      & 1.2
      & $<1.7$
      & $87$ N\\
    J1709$-$1640 & 0.89 & 1640 & 0.56$^\mathrm{a}$ & 125$^\mathrm{d}$ & 6  & 535
      & 1.3
      & $<1.5$
      & $16^{+11}_{-10}$ F\\
    \hline\hline
  \end{tabular}
  \caption{A search for filaments in new observations. \textit{Left}: Pulsar properties (as in table \ref{tab:pulsar}). \textit{Right:} Filament search results (no filaments are found because $\sigma$ is less than 3) A dagger indicates that the feature identified has position angle consistent with a PWN trail. The significance reflecting this alignment is given in parentheses. 0.5--7\,keV unabsorbed point source fluxes are also given in units of $10^{-15}$\,erg\,cm$^{-2}$ s$^{-1}$ (90\% credible intervals, or 90\% upper limits when the measured flux is consistent with zero.). The extended source fluxes are given for structures off the velocity axis (F), proper motion axis  trails (T), and amorphous nebular structures (N). They should be thought of as upper bounds on extended flux from these sources. Distance and proper motion references are $^\mathrm{a}$\citep[DM]{yao2017new}, $^\mathrm{b}$\citep[parallax]{chatterjee2009precision}, $^\mathrm{c}$\citep[parallax]{deller2019microarcsecond}, $^\mathrm{d}$\citep{jankowski2019utmost}, $^\mathrm{e}$\citep{li2016proper}.}
  \label{tab:search}
\end{table*}

\subsection{New Observations: A Snapshot Search for Bright Filaments of Low \texorpdfstring{$r_0$}{Stand-off Distance} Pulsars}
\label{sec:new}

Small bow shock standoff is a salient property of the known filaments. Using the ATNF pulsar catalog \citep{manchester2005australia} to constrain the factors in Eq.~\ref{eqn:r0}, we select targets with similarly low $r_0$ that might have detectable filaments. We consider only pulsars with a detected proper motion ($>3\sigma$ on one or both axes). The ISM density $n_\mathrm{ISM}$ is not directly constrained, but by restricting to pulsars with distance estimate $d<4$ kpc and $|z|<0.2$ kpc, we increase the odds that the pulsars are in the dense warm neutral medium (WNM) or warm ionized medium (WIM) ISM phases. Any one pulsar might lie in the low density hot ionized ISM, giving large $r_0$ and no filament, but this provides a reasonable selection criterion. Ranking by $r_0$, Guitar tops the list, but none of the other 10 smallest $r_0$ had \cxo\ observations. We have obtained $\sim$25 ks ACIS-I snapshots of seven of these targets (see table \ref{tab:search}). Our search method is capable of detecting filaments in such short observations; it detects the five known filaments to $>3 \sigma$ individually in every observation deeper than 20 ks. However, we detect no significant filaments in these new exposures.

\begin{figure}
  \centering
  \includegraphics[width=\linewidth]{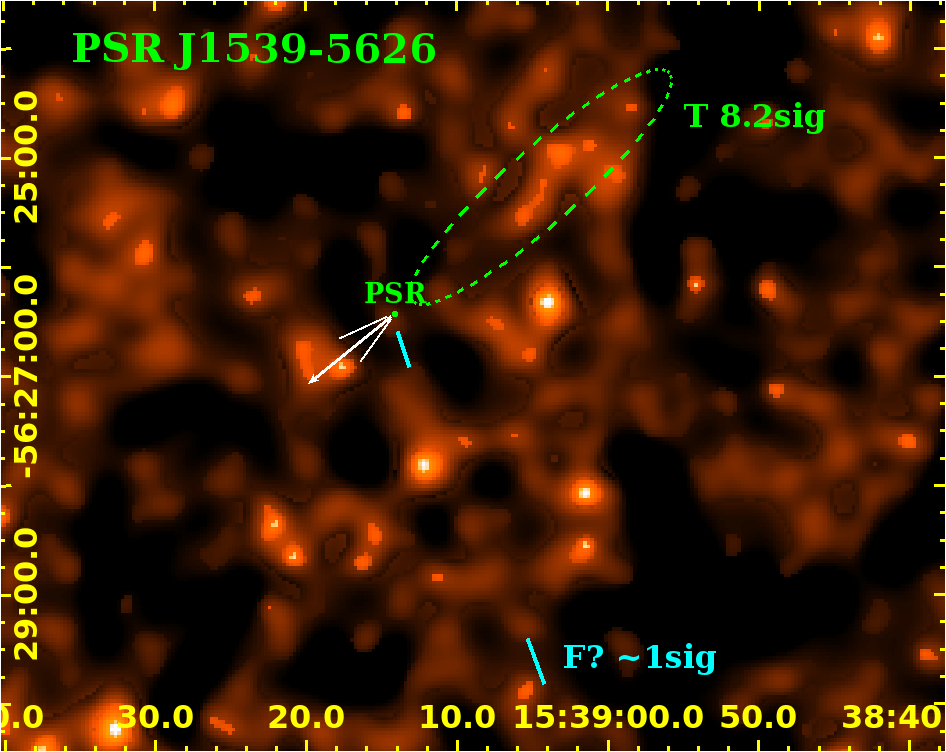}
  \caption{The highest $g$ snapshot survey source produces the best signal from the automatic search; a trail (green) of $8.2\sigma$ significance when searched over the small proper motion range ($10^3$yr length, white). {\it CXO} ACIS $1-7$\,keV adaptively smoothed image. The next best feature found is at large angle (cyan) and is not significant at just over 1$\sigma$. A substantially deeper observation would be required  to map the trail and check the validity of this weak filament candidate.}
  \label{fig:J1539}
\end{figure}

For J1539, we obtain a $1.6\sigma$ detection searching over all directions. However this detection is a trail, anti-aligned with the well-measured pulsar proper motion. Incorporating the alignment within $\pm\sigma_{v_\perp}$ of the trail, the significance rises to $8.2\sigma$, giving this pulsar a good trail detection (Fig.\,\ref{fig:J1539}). This boosted significance is indicated in Fig.~\ref{fig:search-results}. A few other pulsars show boosted significance from alignment with the proper motion as well, including Cannonball (a well known, short X-ray trail) which is boosted to $15 \sigma$ significance. For J1539---which boasts the highest gate fraction of all new pulsars surveyed in this work---the search routine finds a secondary linear feature south of the pulsar. However, at just over 1$\sigma$, the signal is too weak to qualify as a filament detection or bear further analysis.

Like J1539, the filament search identified a $1.5\sigma$ trail-like excess in J1835. This pulsar's less certain proper motion renders this $4.4\sigma$ trail only a candidate. J1835 shows no additional nebulosity.

Table \ref{tab:search} also shows unabsorbed $0.5-7$ keV point source fluxes $F_\mathrm{PS}$ for the seven new observations. These are extracted with the \texttt{ciao} \texttt{srcflux} tool. Fluxes are too low to constrain the point source spectrum, so we assume an absorbed power law with photon index of $\Gamma = 2$ and column density $n_H$ obtained from the $n_H$-DM correlation \citep{he2013correlation}. The point source was detected to $>90$\% confidence in three observations, and upper uncertainty values are presented for the other four. Only four (three) photons were detected within one arcsecond of J1136 (J1539).

Middle age pulsars typically have point-source non-thermal X-ray efficiencies of $\eta_\mathrm{PS}= L_\mathrm{X} / \dot E \approx 10^{-4}$, albeit with substantial dispersion \citep{2022A&A...658A..95V}. Computing efficiencies from the fluxes and distances of table \ref{tab:search}, we find values consistent with this trend: $\eta_\mathrm{PS} = (1.2, 1.6, 1.4) \times 10^{-4}$ for J0454, J1136 and J1539, respectively.  Some extended PWN emission might also be expected, although the PWN efficiency for the relatively old, low ${\dot E}$ pulsars observed here is not well known. Noting that the only clear statistical detection is the trail feature for J1539, we have attempted to identify regions of faint extended emission with some morphological connection to the pulsar position; the fluxes, with all identified point sources removed, are given as $F_\mathrm{ext}$ in Table 4. These should properly be viewed as rough PWN upper limits. However it is interesting to note that the brightest indications of extended flux are for J1539 at total $\eta_\mathrm{ext} \sim 4.2 \times 10^{-3}$ (the most energetic target) and J1833 at $\eta_\mathrm{ext} \sim 1.0 \times 10^{-2}$ (the youngest target).

\subsection{Refined Conditions for Filament Production}
\label{sec:escape}

\begin{figure}
  \centering
  \includegraphics[width=\linewidth]{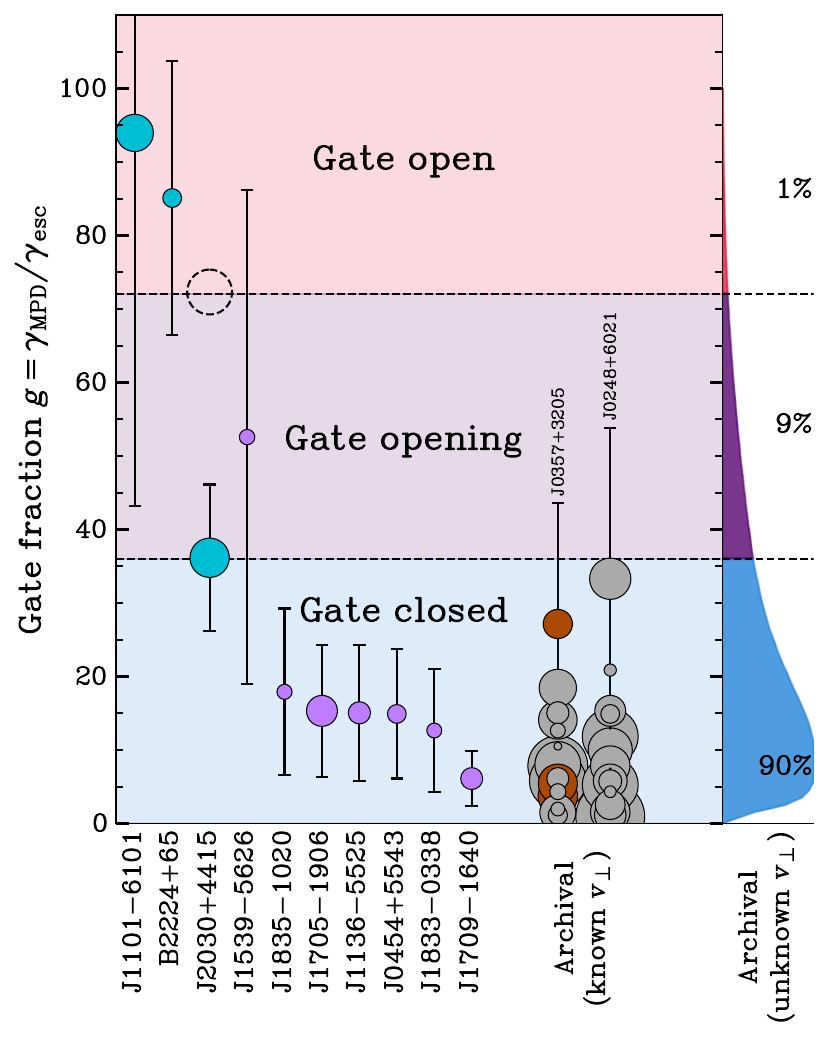}
  \caption{Gate fractions $g$ (Eq.~\ref{eqn:gate-fraction}) of all pulsars discussed in this work. The leftmost three are known filaments (a dotted circle marks the J2030 position with density enhancement), the next seven are newly-observed non-detections, the rightmost two columns are archival survey pulsars with known proper motion, and the rightmost curve represents those with unknown proper motion as a probability distribution. Points have the same colors as in Fig.~\ref{fig:search-results}, radius proportional to $\log(\dot E / d^2)$. The background indicates suggested gate fraction cuts. Error bars indicate combined velocity and $n_\mathrm{ISM}$ uncertainties.}
  \label{fig:gate-fraction}
\end{figure}

With five detected filaments (plus three candidates), seven new non-detections, and the archival non-detections, we can refine the simple model \citep{bandiera2008on} of pulsar properties required for a filament. Focusing on the particle energetics, we consider the Lorentz factor of the most energetic particles
\begin{equation}
  \gamma_\mathrm{MPD} \sim \frac{e}{mc^2}\sqrt{\frac{\dot E}{c}},
  \label{eqn:gamma-mpd}
\end{equation}
the Lorentz factor of particles that can escape onto the filament
\begin{equation}
  \gamma_\mathrm{esc} \sim \frac{eB r_0}{mc^2},
  \label{eqn:gamma-esc}
\end{equation}
and the minimum Lorentz factor required to synchrotron-radiate soft X-rays
\begin{equation}
  \gamma_X \sim \sqrt{\frac{m c E_1}{\hbar B e}}.
  \label{eqn:gamma-x}
\end{equation}
These are obtained from the maximum potential drop (MPD) of the pulsar, the Larmor radius required to equal the stand-off distance, and standard synchrotron theory.

Dimensionless $\gamma$ ratios help us sort pulsars for ability to produce X-ray filaments. If $\gamma_X/\gamma_\mathrm{MPD} > 1$, then no filament can be seen by \cxo. This sets a lower limit on the pulsar $\dot E$ of $\dot E \gtrsim 10^{33}$ erg s$^{-1}$ for $B = 3\ \mu$G. The other ratio is the ``gate fraction''
\begin{equation}
  \begin{aligned}
  g &\equiv \frac{\gamma_\mathrm{MPD}}{\gamma_\mathrm{esc}} = \frac{\sqrt{4\pi \mu m_p n_\mathrm{ISM} v^2_\perp}}{B}\\
  &= 38\parens{\frac{v_\perp}{300\ \mathrm{km}\ \mathrm{s}^{-1}}}\parens{\frac{n_\mathrm{ISM}}{0.5\ \mathrm{cm}^{-3}}}^{1/2}\parens{\frac{B}{3\ \mu\mathrm{G}}}^{-1}.
  \end{aligned}
  \label{eqn:gate-fraction}
\end{equation}
where we have used $v_\perp$ as a more accessible stand-in for $v_\mathrm{psr}$. If $g < g_c$ for some gate fraction cutoff $g_c$, then escape from the bow shock is not possible and the ``gate'' is closed. Geometrical factors, likely traceable to the relative orientation of the ISM $B$, pulsar spin axis, and magnetic inclination should contribute to $g_c$. Lacking a sufficiently detailed filament model, we must estimate it empirically, but we expect $g_c\gg 1$. Pulsars with $g\gg g_c$ will inject a power-law $e^\pm$ spectrum with a modest range of $\gamma$. Pulsars with $g\sim g_c$ should inject nearly mono-energetic $e^\pm$, with injected power growing with with $g-g_c$.

We estimate $g$ for all pulsars discussed in this work. We assume an undisturbed ISM $B$ field of 3 $\mu$G. $n_\mathrm{ISM}$ is computed for the volume dominating ISM phases---WNM, WIM, and hot ionized medium (HIM)---using the analytical estimates for typical density and volume fill factor given as a function of Galactic $|z|$ in \cite{brownsberger2014survey}. We know that the few pulsars with detected H$\alpha$ bow shocks are in the WNM. For those with filaments, we know that they are in the WNM or WIM. For all other pulsars we assume that they randomly sample the ISM, including the low density HIM. In our $g$ estimate we thus use the probability (i.e.~fill factor)-weighted density. When plane-of-sky velocity $v_\perp$ is not well constrained, we use the double-Maxwellian analytical pulsar space velocity distribution of \cite{igoshev2020observed} ($\times \sqrt{2/3}$) as a prior for $v_\perp$ and update the prior with any weak proper motion constraint in the ATNF catalog.

The mean values of this $g$ distribution are listed in tables \ref{tab:pulsar} and \ref{tab:search}. Of course the unknown factors give each $g$ estimate wide uncertainty, so the value for any one pulsar is not very predictive.  Fig.~\ref{fig:gate-fraction} depicts the gate fraction calculated for all pulsars discussed in this work, separating the filaments, newly observed pulsars, and archival pulsars. Error bars incorporate velocity (combined proper motion and an assumed 20\% distance uncertainty) and $n_\mathrm{ISM}$ uncertainties. Filaments exhibit higher gate fractions than all pulsars with detected $>3\sigma$ $v_\perp$ (points). They also are expected to exceed the majority of archival pulsars with unknown proper motion (distribution at right).

Gate fraction appears to be a better filament predictor than stand-off distance. An $r_0$ cutoff selects low-power pulsars, while $g$'s allows large ${\dot E}$. This is because $\dot E$ increases both $r_0$ and $\gamma_\mathrm{MPD}$ equally. Gate fraction is also better correlated with filament luminosity $L_\mathrm{fil}/\dot E$ than $r_0$. The new observations, while outstanding in $r_0$, are more modest in $g$; this renders the non-detections easier to explain with a gate fraction cutoff.

Fig.~\ref{fig:gate-fraction} shows that this cutoff is $g_c\approx 35-70$. If $g_c=35$, J2030 passes the cutoff as do 10\% of the pulsars in our archival survey. However, the multi-bubble nature of J2030's bow shock suggests that the pulsar is passing through an ISM with substantial local density fluctuations, and that during these $g$ will be larger than that of the generic WNM. Indeed, the very narrow J2030 filament suggest that the gate for this pulsar has only recently, and briefly, opened as it passed through such a density enhancement. A dotted circle in Fig.~\ref{fig:gate-fraction} shows the location of J2030 under $4\times$ $n_\mathrm{ISM}$ amplification (as might be expected from past shock compression). This suggests $g_c \approx 70$; all other known-$v_\perp$ pulsars fail this higher cut at their most likely velocities, as do $99$\% of archival pulsars. The only other pulsar in this paper with substantial support for $g>75$ is J1539; however as noted our check for a second (filament) axis gives a low significance at best. It would thus not be surprising if this pulsar had a low $n_{\rm ISM}$ and thus occupied the bottom of its plotted range.

For pulsars near the gate cutoff---e.g.~J2030, J0248, and possibly J1539---it is possible that a faint filament exists and is not detected by \cxo. The relevant quantity for detectability in these cases is the filament flux, which should scale as $\dot E / d^2$ (the size of the points in Fig.~\ref{fig:gate-fraction}). The faintest filament J2030 performs exceptionally well in this metric, while the newly searched pulsars are more distant with lower filament detectability.

\section{Conclusions}
\label{sec:conclusion}

We have measured the observables of the known pulsar X-ray filaments (otherwise termed ``misaligned outflows'') and tabulated these along with the properties of their parent pulsars. We have identified only five secure filaments and three candidates. To further explore the population, we have conducted a uniform survey of the {\it CXO} archive of ACIS pulsar observations. This has been augmented by a 25\,ks/source snapshot survey of seven previously unobserved pulsars selected to have properties most similar to the filament pulsars. Our analysis is sufficiently sensitive to detect at $>3\sigma$ any of the known filaments in exposures of $\ge 20$\,ks, yet we find no secure new filaments and only a handful PSR whose extended emission includes low-significance filament candidates. These survey results confirm that conditions for filament production are special and filaments will be very rare.

Examining the properties of the parent pulsars, we see that a plausible criterion for filament production remains escape of the highest energy PSR/PWN particles from a bow shock apex. We write a simple expression for a ``gate'' parameter $g =\gamma_{\rm MPD}/\gamma_{\rm esc} \propto v_{\rm PSR} n_{\rm ISM}^{1/2} B_{\rm ISM}^{-1}$. Observations support a critical value $g_c \approx 35-70$ which allows such escape. One would like to use this parameter to find additional filament pulsars, but this requires good proper motion measurements and depends on the generally un-measured $n_{\rm ISM}$. Significantly, there is a large overlap between the handful of pulsars producing H$\alpha$ bow shocks and those producing filaments---these bow shocks require a substantially neutral local medium, and hence large $n_{\rm ISM}$. For pulsars with $g\sim g_c$, like J2030, this suggests that the gate can temporarily open as the pulsar passes through local density enhancements, leading to an interment filament. 

Our compilation can assist researchers looking to test analytic and numerical models for filament production. At present only the most basic conditions are evident and we caution that other pulsar parameters (e.g.\, spin and field orientation) may play a critical role in filament production and luminosity. More examples are clearly needed, as well as better measurements of the proper motion, etc.\,for existing filaments, to probe such effects. While our analysis points to pulsar properties that can help select likely filament targets, the challenge of clearing the $g_c$ gate suggests that the search may be long.

\begin{acknowledgments}
This work was supported in part by NASA grant G03-2404X administered by the Smithsonian Astrophysical Observatory.

\end{acknowledgments}

\vspace{5mm}
\facilities{Chandra X-Ray Observatory, Sherlock Computing Cluster (Stanford University)}
\software{\texttt{ciao} \citep{2006SPIE.6270E..1VF}, \texttt{HEAsoft} \cite{2014ascl.soft08004N}, \texttt{Rust}, \texttt{Python}}

\bibliography{bib}{}
\bibliographystyle{aasjournal}

\appendix

\section{Search Method}
\label{app:search}
A search method capable of detecting filaments in low-signal data must deliver accurate estimates of significance $\sigma$ and goodness of fit (GoF) $\Delta$. Our data has few enough counts that the log-likelihood ratio under the null hypothesis (NH) and log-likelihood under the model are not $\chi^2$-distributed as is sometimes the case. In this appendix, we derive the distribution of these two statistics and map them to accurate estimates of $\sigma$ and $\Delta$.

In our filament model, the expected number of background counts in bin $b$ is
\begin{equation}
  \lambda_- = F_0 x_b
\end{equation}
where $x_b$ is the bin exposure generated with the \texttt{fluximage} \texttt{ciao} tool and $F_0$ is the background flux (counts per second per bin). The expected number of filament counts is
\begin{equation}
  \lambda_+ = a S_b F_0 x_b
\end{equation}
where $S_b$ represents a profile for the filament shape (specified by the model) and $a$ represents a unitless filament amplitude. The NH corresponds to $a=0$. We measure $F_0$ from the data by summing the counts in regions where $S_b$ is low, to avoid including filament counts.

The fit is conducted by minimizing likelihood $\mathcal{L}$ (Eq.~\ref{eqn:search-likelihood}) with respect to $a$ and the parameters controlling the filament profile $S_b$ (e.g. the filament width). This is done for every filament orientation from 0 to $360^\circ$ with $1^\circ$ spacing.

\subsection{Significance}
We begin calculating the distribution of $\ln (\mathcal{L}_M / \mathcal{L}_{NH})$ by noting that exposure $x_b$ and the filament profile $S_b$ are either zero or equal to a near-constant nonzero value, which varies slightly over the image. These variations must be captured in $\mathcal{L}$ to avoid bias, but do not substantially affect the distribution of $\mathcal{L}$. We therefore approximate $\lambda_-$ as constant throughout the image, and $\lambda_+ = a \lambda_-$ in some number of bins where the filament is bright, at a level given by
\begin{equation}
  B_\mathrm{eff} = \brackets{\sum_{b=1}^B S_b x_b}\brackets{\sum_{b=1}^B x_b}^{-1}.
\end{equation}
The log-likelihood-ratio is therefore
\begin{equation}
  \begin{aligned}
  \ln(\mathcal{L}_M / \mathcal{L}_{NH}) = &\sum_{b=1}^{B_\mathrm{eff}} \Bigg[\parens{n_b \ln (1 + a) - a\lambda_-}\\
  &+ \sum_{i=1}^{n_b}\ln\parens{\frac{P_-(E) + a P_+(E)}{P_-(E) + a P_-(E)}}\Bigg].
  \end{aligned}
\end{equation}

The last term is generally small because $P_+(E)$ is usually small under the NH. We therefore neglect it. In the first line, the only bin-dependent term is $n_b$, which can be explicitly summed over by introducing $N = \sum_{b=1}^{B_\mathrm{eff}} n_b$. The best fit value of $a$ is also related to $N$ by $N = (1 + a)\lambda_-$. Thus, the log-likelihood ratio simplifies to
\begin{equation}
  \ln(\mathcal{L}_M / \mathcal{L}_{NH}) \sim N\ln \parens{\frac{N}{\lambda_-}} - N + \lambda_-.
  \label{eqn:likelihood-ratio-simplified}
\end{equation}
The cumulative distribution function (CDF) of $N$ is Poissonian with expected count number $\lambda_- B_\mathrm{eff}$. The CDF of the log-likelihood ratio $C_1[\ln(\mathcal{L}_M / \mathcal{L}_{NH})]$ for a single fit is determined by numerically inverting Eq.~\ref{eqn:likelihood-ratio-simplified}.

Since we perform $A$ fits at different angles and report the best overall log-likelihood ratio, the true CDF of this ratio is elevated to $C_1[\ln(\mathcal{L}_M / \mathcal{L}_{NH})]^A$. We run $360$ fits, but since filaments offset by one degree overlap spatially, the amplitudes for adjacent fits are correlated and the effective number of independent fits is fewer than 360. Running the analysis on simulated data sets for multiple values of $\lambda_-$ reveals that $A \approx 75$ is accurate for the number of bins used in the archival search. Defining
\begin{equation}
  \sigma = \sqrt{2} \mathrm{erf}\parens{1 - 2C_1\parens{\ln\frac{\mathcal{L}_M}{\mathcal{L}_{NH}}}^{75}}
\end{equation}
where $\mathrm{erf}$ is the error function 
guarantees that $\sigma$ is standard-Gaussian-distributed under the NH.

\subsection{Goodness of Fit}

The distribution of the log-likelihood ratio was not $\chi^2$ distributed because it reduced to a single parameter $N$. However, the likelihood itself is a function of the fluxes in many bins and is Gaussian distributed by the central limit theorem. Defining
\begin{equation} 
  \Delta = \frac{\mathcal{L} - \langle \mathcal{L} \rangle}{\mathrm{Var}\ \mathcal{L}},
  \label{eqn:delta}
\end{equation}
therefore ensures that $\Delta$ is distributed according to the standard Gaussian. Here, $\langle \mathcal{L} \rangle$ and $\mathrm{Var}\ \mathcal{L}$ are the expected likelihood and likelihood variance under the model.

The Poisson part of the likelihood has expected value
\begin{equation}
    \sum_{b=1}^B \lambda \ln \lambda - \lambda - \langle n_b! \rangle
\end{equation}
and variance
\begin{equation}
    \sum_{b=1}^B \lambda (\ln \lambda)^2 + \mathrm{Var}(\ln n_b!) - 2 \ln \lambda \mathrm{Cov}(n_b, \ln n_b!).
\end{equation}
The variance and covariance are integrated numerically.

The energy part of the likelihood has expected value
\begin{equation}
    \sum_{b=1}^B \lambda_+ \langle X \rangle_+ + \lambda_- \langle X \rangle_-\\
\end{equation}
and variance
\begin{equation}
  \begin{aligned}
  \sum_{b=1}^B &(\lambda_+ + \lambda_+^2) \langle X^2 \rangle_+ + (\lambda_- + \lambda_-^2)\langle X^2 \rangle_-\\
   &- (\lambda_+ \langle X \rangle_+)^2 - (\lambda_- \langle X \rangle_-)^2\\
  \end{aligned}
\end{equation}
where
\begin{equation}
  X = \ln \frac{P_-(E) + a P_+(E)}{1+a}.
\end{equation}
$+$ and $-$ subscripts on the expected values indicate whether the average is taken under the $P_+$ or $P_-$ distribution. The expected values of $X$ and $X^2$ cannot be computed analytically, so we randomly generate 1,000 samples from each spectrum and numerically evaluate the distribution of $X$. Neglecting the small covariance between the energy term and the Poisson term, we may simply add these results to compute the expected value and variance given in Eq.~\ref{eqn:delta}.

Though this value of $\Delta$ can be computed for all bins in the image, we only compute it for bins where $S_b$ is large, because we are more interested in the deviation between the \textit{filament} data from the model rather than the background noise.

The distributions of $\sigma$ and $\Delta$ were both confirmed to be standard Gaussians by running our analysis on many mock \cxo\ observations. These were simulated by Poisson-drawing counts from sample \cxo\ exposure maps and likewise drawing event energies from sample spectra.

\section{Equipartition Expression}
\label{app:equipartition}

Our result for the equipartition magnetic field (Eq.~\ref{eqn:equipartition}) differs from one in common use in the literature \citep[e.g.][]{pavlov2003variable} due to the presence of $X(\Gamma)$. For $\gamma \sim 1-2$, the two formulas lie within 25\% of each other, while the difference grows to a factor of almost 2 for $\Gamma = 3$. Our field is weaker and depends less strongly on $\Gamma$. It therefore functions as a more conservative estimate of $B_\mathrm{eq}$ in this use case.

Our formula is obtained by minimizing the total energy density $U_B + U_K$ with respect to $B$, where $U_B = B^2 / 8\pi$ is the ISM magnetic energy density and $U_K$ is the kinetic energy density of relativistic particles. This minimum occurs at $U_B / U_K = 3/4$. These particles are assumed to be power-law distributed, with isotropically distributed pitch angle (which we average over). The connection between the normalization of this power law and the source luminosity are both given by standard synchrotron theory \citep[Eq.~6.36 of][]{rybicki1991radiative}; the $X(\Gamma)$ term arises from this relation.

\section{Point source subtraction method}
\label{app:star}

The archival and spectral analysis of this work require excising point sources from the CXO observations. The archival analysis in particular is sensitive to uncleaned point sources, and the shallowness of many archival observations renders typical point sources only tenuously detected. Thus aggressive cuts are required.

We detect point sources by fitting a Gaussian-plus-background radial surface brightness profile SB$(r)=a \exp[-r^2/(2s^2)] + b$ to the image near each count. Events where the Gaussian term $a$ is is significantly elevated above background $b$ and where $s$ is comparable to the local PSF width are treated as point-source candidates. We approximate the PSF width $s_\mathrm{PSF}$ as quadratically increasing with target location from 0.4$''$ on-axis to 2.5$''$ at $6'$ off-axis. Point source detection is performed on stacked observations, making the method particularly sensitive for deep observations such as those of the known filaments (in which 10-20\% of the counts within the region of interest are typically removed).

For morphological analyses, we remove all events associated with point sources. Good star mask conditions were found to be $s < 3 s_\mathrm{PSF}$ and $(a-b)/\sigma_a > 1$\footnote{The latter condition should not be regarded as an exact statistical statement. The $a$ uncertainty is estimated using $\chi^2$ fitting which is not statistically valid for the low counts present here, but still serves as a useful metric for point source removal.}. For spectral analyses, we are studying the statistical behavior of a larger set of counts. We then use more relaxed cuts of $s < 2.5 s_\mathrm{PSF}$ and $(a-b) / \sigma_a > 1.5$ and de-weight each event by the local amplitude of the Gaussian-component of the point source model rather than cutting the event completely.

\end{document}